\documentclass[11pt,a4paper]{article}

\usepackage{colortbl}
\usepackage{ascmac}
\usepackage{bm}
\usepackage[dvipdfmx]{graphicx}
\usepackage{amsmath}
\usepackage{amsfonts}

\usepackage{youngtab}

\def\tr{\mathop{\rm tr}\nolimits}
\def\Pexp{\mathop{\rm Pexp}\nolimits}

\begin{document}

%%%%%%%%%%%%%%%%%%%%%%%%%%%%%%%%%%%%%%%%
\Yautoscale0
\newcommand{\YT}{\Yboxdim15pt\young}
\newcommand{\y}{\Yboxdim4pt\yng}
\newcommand{\Y}{\Yboxdim6pt\yng}
\Yvcentermath1
\newcommand{\R}{\rotatebox[origin=b]{90}}
%%%%%%%%%%%%%%%%%%%%%%%%%%%%%%%%%%%%%%%%

%titlepage
\begin{titlepage}
\title{
\vspace{-1.5cm}
\begin{flushright}
{\normalsize TIT/HEP-702\\ June 2024}
\end{flushright}
\vspace{1.5cm}
\LARGE{Giant graviton expansion for general Wilson line operator indices}}
\author{
Yosuke {\scshape Imamura$^1$\footnote{E-mail: imamura@phys.titech.ac.jp}},
Akihiro {\scshape Sei$^1$\footnote{E-mail: a.sei@th.phys.titech.ac.jp}},
and
Daisuke {\scshape Yokoyama$^2$\footnote{E-mail: ddyokoyama@meiji.ac.jp}} \\
\\[18pt]
{\itshape $^1$Department of Physics, Tokyo Institute of Technology}, \\ {\itshape 2-12-1 Ookayama, Meguro-ku, Tokyo, Japan} \\
% {\itshape and} \\
\\[-12pt]
{\itshape $^2$Department of Physics, Meiji University}, \\ {\itshape 1-1-1 Higashi-Mita, Tama-ku, Kawasaki-shi, Kanagawa, Japan}
}

\date{}
%\date{July 12, 2019}
\maketitle
\thispagestyle{empty}
\begin{abstract}
We propose a giant graviton expansion for Wilson line operator indices
in general representations.
The inserted line operators are specified by power sum symmetric polynomials $p_\lambda$
labeled by partitions $\lambda$.
We interpret the partitions as the structure of fundamental string worldsheets wrapping around the temporal circle.
The strings may or may not end on giant gravitons,
and by summing the contributions from all brane configurations
consistent with the specified partitions,
we obtain the finite $N$ line operator index.
The proposed formula is consistent with known results
and passes highly non-trivial numerical tests.
\end{abstract}

\end{titlepage}

\tableofcontents

\section{Introduction}
The superconformal index \cite{Romelsberger:2005eg,Kinney:2005ej} is a useful tool
to study supersymmetric field theories.
We can exactly calculate the index for a large class of field theories
and it enables us to analyze non-perturbative aspects such as dualities
and the AdS/CFT correspondence \cite{Maldacena:1997re,Gubser:1998bc,Witten:1998qj}.

It was found in the seminal work \cite{Kinney:2005ej} that the large $N$ index of the
${\cal N}=4$ SYM
is reproduced by the index of the supergravity multiplet
in $AdS_5\times S^5$.
Furthermore, we can calculate finite $N$ corrections to the index on the gravity side
as the contributions
from giant gravitons \cite{McGreevy:2000cw,Mikhailov:2000ya} with different wrapping numbers.
Namely, the finite $N$ index can be expressed as an expansion over wrapping numbers
of giants.
Such expansions are called giant graviton expansions
\cite{Arai:2019xmp,Arai:2020qaj,Imamura:2021ytr,Gaiotto:2021xce}.

Recently, giant graviton expansions were extended to the index with Wilson line operator insertions.
In \cite{Imamura:2024lkw} the insertion of Wilson line operators
in the fundamental and the anti-fundamental
representations was studied,
and in \cite{Imamura:2024pgp} a giant graviton expansion for
the anti-symmetric representations was proposed.
See \cite{Beccaria:2024oif,Beccaria:2024dxi} for approaches from the gauge theory side.
See also \cite{Gaiotto:2021xce,Kim:2024ucf} for giant graviton expansions of the index with surface operator insertions,
and \cite{Hatsuda:2024uwt} for a mutual relation between half indices and line operator indices.
The purpose of this work is to generalize the expansion in \cite{Imamura:2024lkw}
for the fundamental representation to lines in general representations.

We consider ${\cal N}=4$ $U(N)$ SYM and define
the six Cartan generators
of the superconformal algebra $psu(2,2|4)$:
\begin{align}
H,\quad
J_1,\quad
J_2,\quad
R_x,\quad
R_y,\quad
R_z,
\end{align}
where $H$ is the Hamiltonian,
$J_1$ and $J_2$ are spin operators, and $R_x$, $R_y$, and $R_z$ are Cartan generators
of the $SU(4)_R$ symmetry.
Wilson line operator indices \cite{Dimofte:2011py,Gang:2012yr,Drukker:2015spa,Hatsuda:2023iwi,Guo:2023mkn,Hatsuda:2023imp}
are defined by operator insertion
from the Schur index \cite{Gadde:2011uv}, which
is a specialization of the superconformal index.
We define the Schur index by
\begin{align}
I=\tr((-1)^Fq^{J_1}x^{R_x}y^{R_y}),\quad
q=xy.
\label{Schur}
\end{align}
See \cite{Bourdier:2015wda,Pan:2021mrw,Hatsuda:2022xdv} for
analytic formulas of the Schur index without line insertions.

We can calculate the Schur index as the partition function of the theory
in $S^3\times S^1$,
and line operator indices as
correlators of line operators.
We insert two line operators at two antipodal points in $S^3$
so that the insertion does not break symmetries used in the definition of the Schur index (\ref{Schur}).
The path integral reduces to the following holonomy integral
\begin{align}
\langle{\cal O}(U){\cal O}'(U^{-1})\rangle^{U(N)}
=\int_{U(N)}\hspace{-1.3em} dU
\Pexp(f^{({\rm vec})}\chi_{\rm adj}(U))
{\cal O}(U){\cal O}'(U^{-1}),
\label{localization}
\end{align}
where $\int_{U(N)}dU$ is the integral over the gauge group $U(N)$ with the normalized Haar measure,
and $\Pexp$ is the plethystic exponential.
The letter index is the product of the
letter index of the ${\cal N}=4$ vector multiplet
\begin{align}
f^{({\rm vec})}=1-\frac{(1-x)(1-y)}{1-q},
\end{align}
and
the character of the $U(N)$ adjoint representation $\chi_{\rm adj}(U)$.
${\cal O}(U)$ and ${\cal O}'(U^{-1})$ are
gauge invariant functions of the holonomy $U$ associated with
the two lines.
We usually diagonalize $U$ in practical calculations, and treat ${\cal O}(U)$ and ${\cal O}'(U^{-1})$ as
symmetric functions of $N$ diagonal components of $U$.

%%%%%%%%%%%%%%%%%%%%%%%%%%%%%%%%%%%%%%%%%%%%%%
\paragraph{Large $N$ index}
Let us review the
line operator indices in the large $N$ limit
\cite{Gang:2012yr,Hatsuda:2023imp,Hatsuda:2023iof}.

A natural basis of Wilson line operators is the set of ones associated with the $U(N)$ irreducible representations,
which are labeled by partitions $\mu$.
In the localization formula (\ref{localization}), they appear as the characters of the representations.
The characters of $U(N)$ irreducible representations $\mu$ are called the Schur polynomials,
and denoted by $s_\mu(U)$.
For the following analysis on the AdS side, however,
as was pointed out in \cite{Hatsuda:2023imp},
it is more convenient to use the power sum symmetric polynomials $p_\lambda(U)$ labeled by partitions $\lambda$:
\begin{align}
p_\lambda(U)=\tr[P_\lambda U^{\otimes|\lambda|}]=\prod_{i=1}^{\ell(\lambda)}(\tr U^{\lambda_i}),
\label{tlambda}
\end{align}
where $\ell(\lambda)$ is the number of parts in $\lambda$ and $|\lambda|$ is the total of $\lambda$.
$p_\lambda(U)$ are related to the Schur polynomials
by Frobenius formula
\begin{align}
p_\lambda(U)=\sum_{\mu,\ell(\mu)\leq N} \chi_\mu (\lambda) s_\mu(U),
\label{frobenius}
\end{align}
where $\chi_\mu(\lambda)$ are the characters of a representation
of the symmetry group $S_k$
labeled by a partition $\mu$ evaluated at a conjugacy class specified by $\lambda$.
$U^{\otimes k}$ in (\ref{tlambda}) acts on the tensor product of $k$ copies of $N$-dimensional vector space $V$,
and $P_\lambda$ is the permutation of the vector spaces specified by $\lambda$.

It is convenient to introduce the notation
\begin{align}
[f(x)]_\lambda=\prod_{i=1}^{\ell(\lambda)} f(x^{\lambda_i}).
\label{bracket}
\end{align}
By definition, the following relation holds
\begin{align}
[f(x)]_{\lambda}[f(x)]_{\lambda'}=[f(x)]_{\lambda+\lambda'},
\label{additive}
\end{align}
where $\lambda+\lambda'$ is the union of partitions $\lambda$ and $\lambda'$,
which is a partition of $|\lambda|+|\lambda'|$.
$p_\lambda$ can be expressed as $p_\lambda(U)=[\tr U]_\lambda$.

\begin{figure}[htb]
\centering
\includegraphics[scale=0.7]{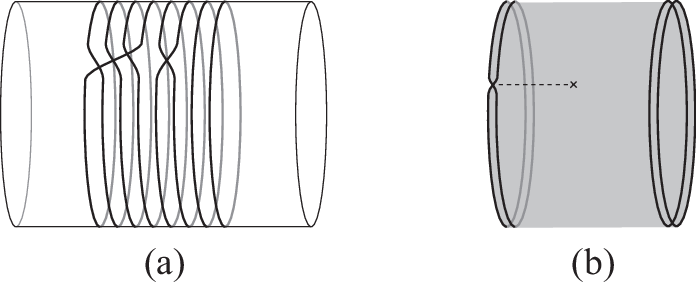}
\caption{(a) World lines $B_\lambda$ with $\lambda=\{4,2,1,1\}$ are shown.
(b) An example of a worldsheet with the Chan-Paton structures on two sides are different.}\label{worldline.eps}
\end{figure}

With the notation (\ref{bracket}) we can rewrite the plethystic exponential as
\begin{align}
\Pexp(f(x))=\sum_\lambda\frac{1}{z_\lambda}[f(x)]_\lambda.
\label{pexplambda}
\end{align}
$z_\lambda$ is defined by $z_\lambda=\prod_{m=1}^\infty m^{r_m}r_m!$,
where $r_m$ is the number of occurrences of $m$ in $\lambda$.
Let us regard $f(x)$ in (\ref{pexplambda}) as a letter index of a particle,
and $\Pexp(f(x))$ as the corresponding full index, which is
the sum of contributions
from different particle numbers $n=0,1,2,\ldots$.
For each $n$ we can depict particles as a bundle of $n$ worldlines,
and when they go around the temporal circle,
they can connect in different ways.
The structure of worldlines, which we refer to as the Chan-Paton structure, can be specified with
partitions $\lambda$ with $|\lambda|=n$.
Let $B_\lambda$ be the bundle of worldlines specified by
$\lambda=\{\lambda_1,\lambda_2,\ldots,\lambda_{\ell(\lambda)}\}$.
It consists of $\ell(\lambda)$ components, and each
component is a loop wrapping $\lambda_i$ times around
the temporal circle.
(Figure \ref{worldline.eps} (a).)
Each loop gives the factor $f(x^{\lambda_i})$,
and together with the statistical factor $1/z_\lambda$,
the contribution from a bundle $B_\lambda$ is given by
the summand in (\ref{pexplambda}).
By summing up contributions from all $\lambda$,
we obtain the full index  (\ref{pexplambda}).

A nice property of the line operators associated with the functions $p_\lambda$
is that they form an orthogonal basis in the large $N$ limit in the sense that
$\langle p_\lambda(U)p_{\lambda'}(U^{-1})\rangle^{U(\infty)}\propto \delta_{\lambda,\lambda'}$.
This can be easily shown by using
character expansion method \cite{Dolan:2007rq,Dutta:2007ws};
\begin{align}
I_{N,\lambda,\lambda'}
&\equiv \langle p_\lambda(U) p_{\lambda'}(U^{-1})\rangle^{U(N)}
\nonumber\\
&=\int_{U(N)}\hspace{-1.3em} dU \sum_{\lambda''}\frac{1}{z_{\lambda''}}f^{({\rm vec})}_{\lambda''}(x)p_{\lambda''+\lambda}(U)p_{\lambda''+\lambda'}(U^{-1})
\nonumber\\
&=\int_{U(N)}\hspace{-1.3em} dU \sum_{\lambda''}\frac{1}{z_{\lambda''}}f^{({\rm vec})}_{\lambda''}(x)
\sum_{\mu,\ell(\mu)\leq N} \chi_\mu(\lambda''+\lambda)s_\mu(U)
\sum_{\nu,\ell(\nu)\leq N} \chi_\nu(\lambda''+\lambda')s_\nu(U^{-1})
\nonumber\\
&=\sum_{\lambda''}\frac{1}{z_{\lambda''}}f^{({\rm vec})}_{\lambda''}(x)
\sum_{\mu,\ell(\mu)\leq N} \chi_\mu(\lambda''+\lambda)\chi_\mu(\lambda''+\lambda').
\end{align}
We used (\ref{localization}), (\ref{additive}), and (\ref{pexplambda}) at the second equality
and (\ref{frobenius}) at the third equality.
At the last equality we used the orthonormality
\begin{align}
\int_{U(N)}\hspace{-1.3em} dU s_\mu(U)s_\nu(U^{-1})=\delta_{\mu,\nu}.
\end{align}
In the large $N$ limit, the bound $\ell(\mu)\leq N$ in the summation over $\mu$ is removed,
and we can use another orthogonality
\begin{align}
\sum_\mu \chi_\mu(\lambda)\chi_\mu(\lambda')=z_\lambda \delta_{\lambda,\lambda'}.
\end{align}
Then, we obtain
\begin{align}
I_{\infty,\lambda,\lambda'}
&=\delta_{\lambda,\lambda'}\sum_{\lambda''}\frac{z_{\lambda''+\lambda}}{z_{\lambda''}}f^{({\rm vec})}_{\lambda''}(x)
\nonumber\\
&=z_\lambda\delta_{\lambda,\lambda'}\prod_{m=1}^\infty\frac{1}{(1-f^{({\rm vec})}(x^m))^{r_m+1}}
\nonumber\\
&=z_\lambda \delta_{\lambda,\lambda'} [I_{\rm F1}]_\lambda I_{\rm sugra},
\label{largen}
\end{align}
where $I_{\rm sugra}$ and $I_{\rm F1}$ are
the index of the gravity multiplet in the AdS bulk \cite{Kinney:2005ej}
and that of fluctuation modes on a single string worldsheet
extended along $AdS_2\subset AdS_5$ obtained in \cite{Gang:2012yr}
based on the mode analysis in \cite{Drukker:2000ep,Faraggi:2011bb}:
\begin{align}
I_{\rm sugra}&=\Pexp\left(\frac{x}{1-x}+\frac{y}{1-y}-\frac{q}{1-q}\right),\nonumber\\
I_{\rm F1}&=\Pexp(x+y-q)=\frac{1-q}{(1-x)(1-y)}=1+\frac{x}{1-x}+\frac{y}{1-y}.
\end{align}

The large $N$ result
(\ref{largen})
can be interpreted on the AdS side as follows \cite{Hatsuda:2023iwi}.
The objects corresponding to the inserted line operators $p_\lambda(U)$ and $p_{\lambda'}(U^{-1})$
are strings extending along $AdS_2\subset AdS_5$.
The bundles of worldlines $B_\lambda$ and $B_{\lambda'}$ are identified with boundaries of the worldsheets \cite{Rey:1998ik,Maldacena:1998im}.
Let us consider a stack of $|\lambda|$ semi-infinite strings
ending on $B_\lambda$ and another
stack of $|\lambda'|$ semi-infinite strings ending on $B_{\lambda'}$.
The two stacks must be connected in the bulk,
and it is possible only when $|\lambda|=|\lambda'|$.
In addition, even if this condition is satisfied,
if $\lambda$ and $\lambda'$ are not the same,
we need to introduce branch points on the  worldsheets.
(Figure \ref{worldline.eps} (b).)
Such worldsheets have higher genus and the amplitude depends
on the string coupling constant $g_{\rm str}$.
We expect such configurations cannot contribute to the index,
and hence we impose the constraint $\lambda=\lambda'$.
$\delta_{\lambda,\lambda'}$ in (\ref{largen}) expresses this constraint.
The combinatorial factor $z_\lambda$ is the number of ways of matching between
two stacks of strings.
The structure of the string worldsheets around the temporal circle
is again specified by $\lambda$($=\lambda'$), and the
index of the fluctuation modes on the strings is given by $[I_{\rm F1}]_\lambda$.

We remark that for finite $N$ the orthogonality of the line operators breaks,
and $I_{N,\lambda,\lambda'}$ are non-vanishing
even when $\lambda\neq\lambda'$ as long as $|\lambda|=|\lambda'|$.
In the following we see that the finite $N$ indices $I_{N,\lambda,\lambda'}$, including the non-diagonal ones,
are reproduced by taking account of contributions of giant gravitons.

%%%%%%%%%%%%%%%%%%%%%%%%%%%%%%%%%%%%%%%%%%%%%%%%%%%%%
\section{Double-sum expansion}\label{doublesum.sec}
The giant graviton expansion for the Schur index without operator insertion
is \cite{Arai:2020qaj}:
\begin{align}
\frac{I_N}{I_{\rm sugra}}
&=\sum_{m_x,m_y=0}^\infty
x^{m_xN}
y^{m_yN}
(xy)^{m_x m_y}
F^{(x)}_{m_x}
F^{(y)}_{m_y}.
\label{noinsertion}
\end{align}
$m_x$ and $m_y$ are the wrapping numbers of giants
wrapped around $C_x$ and $C_y$, respectively,
where $C_x$ and $C_y$ are the three spheres defined as the $R_x$-fixed locus and the $R_y$-fixed locus
in the internal space $S^5$.
The factors $x^{m_xN}$ and 
$y^{m_yN}$ are the classical contributions
from giants.
$F^{(x)}_{m_x}$ is the index of the $U(m_x)$ gauge theory realized on $m_x$ giants
wrapped on $C_x$.
$F^{(y)}_{m_y}$ is defined similarly.
They are related to the Schur index $I_m$ of the ${\cal N}=4$ $U(m)$ SYM
by \cite{Arai:2020qaj}
\begin{align}
F^{(x)}_m=\sigma_x I_m,\quad
F^{(y)}_m=\sigma_y I_m,
\end{align}
where $\sigma_x$ and $\sigma_y$ are the variable changes
\begin{align}
\sigma_x:(q,x,y)\rightarrow(y,x^{-1},q),\quad
\sigma_y:(q,x,y)\rightarrow(x,q,y^{-1}).
\end{align}

In AdS/CFT correspondence,
a fundamental Wilson line operator is realized
as the boundary of a fundamental string worldsheet \cite{Rey:1998ik,Maldacena:1998im}.
Corresponding to the line operators inserted at antipodal points in $S^3$,
we introduce two semi-infinite strings.
By connecting them, we obtain a single worldsheet extended on $AdS_2\subset AdS_5$.

If $N$ is finite and giant gravitons are present,
the semi-infinite strings can end on the giants.
The giant graviton expansion for fundamental lines with $\lambda=\lambda'=\Y(1)$
is \cite{Imamura:2024lkw}
\begin{align}
\frac{I_{N,\y(1),\y(1)}}{I_{\rm sugra}}
&=\sum_{m_x,m_y=0}^\infty
x^{m_xN}
y^{m_yN}
(xy)^{m_x m_y}
\nonumber\\&
\left[
I_{\rm F1}
F^{(x)}_{m_x}
F^{(y)}_{m_y}
+
\frac{1}{x}(I_{\rm F1}^{(x)})^2F^{(x)}_{m_x,\y(1),\y(1)}
F^{(y)}_{m_y}
+
\frac{1}{y}
(I_{\rm F1}^{(y)})^2
F^{(x)}_{m_x}
F^{(y)}_{m_y,\y(1),\y(1)}
\right].
\label{ggefund}
\end{align}
In (\ref{ggefund})
the contribution from specific wrapping numbers $(m_x,m_y)$ consists of
three terms
corresponding to three ways of connection of the two
semi-infinite fundamental strings:
the direct connection,
the connection via giants on $C_x$,
and the connection via giants on $C_y$.

$I_{\rm F1}^{(x/y)}$ is the index of fluctuation modes on a single semi-infinite
string ending on a giant on $C_{x/y}$ \cite{Imamura:2024lkw}:
\begin{align}
&I_{\rm F1}^{(x)}=\Pexp(y-xy),\quad
I_{\rm F1}^{(y)}=\Pexp(x-xy).
\end{align}

From the viewpoint of theories realized on giants,
strings ending on the giants can be treated as line operator insertions.
$F^{(x/y)}_{m,\lambda,\lambda'}$ are line-operator
indices of the theory on a stack of $m$ giants on $C_{x/y}$.
They are given by
\begin{align}
&F^{(x)}_{m,\lambda,\lambda'}=\sigma_x I_{m,\lambda,\lambda'},\quad
F^{(y)}_{m,\lambda,\lambda'}=\sigma_y I_{m,\lambda,\lambda'}.
\end{align}
For $m=0$, the functions $F_{0,\lambda,\lambda'}^{(x/y)}$ are zero
except for $F_{0,\cdot,\cdot}^{(x/y)}=F_0^{(x/y)}=1$.
See Appendix \ref{app.sec} for explicit forms of these functions.

The factors $1/x$ and $1/y$ in the second and the third terms in (\ref{ggefund})
were first introduced in \cite{Imamura:2024lkw} by hand
to reproduce the correct answer.
It was pointed out in \cite{Beccaria:2024lbt} that
they are reproduced as the contribution from zero-point charges.

A lesson from
(\ref{ggefund})
is that we obtain finite $N$ line-operator indices
by summing up contributions of all ways of connection
of semi-infinite strings.
With this lesson in mind,
let us generalize (\ref{ggefund}) to
more general line operator indices $I_{N,\lambda,\lambda'}$.

We want an expansion in the form
\begin{align}
\frac{I_{N,\lambda,\lambda'}}{I_{\rm sugra}}
&=\sum_{m_x,m_y=0}^\infty S_{N,\lambda,\lambda'}^{(m_x,m_y)},
\label{gge0}
\end{align}
where each function $S_{N,\lambda,\lambda'}^{(m_x,m_y)}$ is
the contribution from the system consisting of
giants with wrapping numbers $(m_x,m_y)$
and two stacks of semi-infinite strings with
the Chan-Paton structures specified by $\lambda$ and $\lambda'$.
There are different ways of connection of two stacks of strings,
and as the sum of their contributions
we propose the formula
\begin{align}
S_{N,\lambda,\lambda'}^{(m_x,m_y)}
&=(-1)^{\ell(\lambda)+\ell(\lambda')}
\sum_{\lambda=\lambda_0+\lambda_x+\lambda_y}
\sum_{\lambda'=\lambda'_0+\lambda'_x+\lambda'_y}
z_{\lambda_0}\delta_{\lambda_0,\lambda'_0}[I_{\rm F1}]_{\lambda_0}
\nonumber\\&
\times x^{m_xN}\frac{1}{x^{|\lambda_x|}}F^{(x)}_{m_x,\lambda_x,\lambda'_x}[I_{\rm F1}^{(x)}]_{\lambda_x+\lambda'_x}
\nonumber\\&
\times y^{m_yN}\frac{1}{y^{|\lambda_y|}}F^{(y)}_{m_y,\lambda_y,\lambda'_y}[I_{\rm F1}^{(y)}]_{\lambda_y+\lambda'_y}
\nonumber\\&
\times(xy)^{m_x m_y}.
\label{gge}
\end{align}
$\sum_{\lambda=\lambda_0+\lambda_x+\lambda_y}$ is the summation over all decompositions of
$\lambda$ into three partitions $\lambda_0$, $\lambda_x$, and $\lambda_y$.
A decomposition of $\lambda$ into three partitions is achieved by assigning each part $\lambda_i$ to one of
$\lambda_0$, $\lambda_x$, and $\lambda_y$.
There are $3^{\ell(\lambda)}$ ways of decompositions.
$\lambda_0$, $\lambda_x$ and $\lambda_y$ may be empty.
The same decomposition may occurs multiple times, and then they are taken separately into account.
$\sum_{\lambda'=\lambda'_0+\lambda'_x+\lambda'_y}$ is defined in the same way.

For each sum over decompositions, there are $3^{\ell(\lambda)}$ (or $3^{\ell(\lambda')}$) summands.
Among $3^{\ell(\lambda)+\ell(\lambda')}$ summands of the double sum,
only ones satisfying $\lambda_0=\lambda'_0$, $|\lambda_x|=|\lambda'_x|$, and $|\lambda_y|=|\lambda'_y|$ are non-vanishing.
See Table \ref{decompositions} for an example.

$1/x^{|\lambda_x|}$ and $1/y^{|\lambda_y|}$ are the mismatch factors
associated with strings ending on $C_x$ and $C_y$, respectively.

\begin{table}[htb]
\caption{
List of decompositions giving non-trivial contributions for
$\lambda=\protect\y(2,1)$ and
$\lambda'=\protect\y(1,1,1)$.
The third column shows the multiplicities.}\label{decompositions}
\centering
\begin{tabular}{ccccc}
\hline
\hline
$(\lambda_0,\lambda_x,\lambda_y)$ & $(\lambda'_0,\lambda'_x,\lambda'_y)$ & mult. \\
\hline
$(\cdot,\cdot,\y(2,1))$ & $(\cdot,\cdot,\y(1,1,1))$ & $1$ \\
$(\cdot,\y(2),\y(1))$ & $(\cdot,\y(1,1),\y(1))$ & $3$ \\
$(\cdot,\y(1),\y(2))$ & $(\cdot,\y(1),\y(1,1))$ & $3$ \\
$(\y(1),\cdot,\y(2))$ & $(\y(1),\cdot,\y(1,1))$ & $3$ \\
$(\cdot,\y(2,1),\cdot)$ & $(\cdot,\y(1,1,1),\cdot)$ & $1$ \\
$(\y(1),\y(2),\cdot)$ & $(\y(1),\y(1,1),\cdot)$ & $3$ \\
\hline
\end{tabular}
\end{table}

We found the necessity of
the overall sign factor
$(-1)^{\ell(\lambda)+\ell(\lambda')}$
by comparison to the results on the gauge theory side.
The necessity of this factor implies that the
ground state of a semi-infinite string should be treated as
a fermionic state.
Indeed, the sign factor can be combined with
the factors from semi-infinite strings as
\begin{align}
(-1)^{\ell(\lambda)+\ell(\lambda')}
[I_{\rm F1}^{(x)}]_{\lambda_x+\lambda'_x}
[I_{\rm F1}^{(y)}]_{\lambda_y+\lambda'_y}
=
[-I_{\rm F1}^{(x)}]_{\lambda_x+\lambda'_x}
[-I_{\rm F1}^{(y)}]_{\lambda_y+\lambda'_y}
\end{align}
under the condition $\lambda_0=\lambda'_0$.

For the trivial line insertion $\lambda=\lambda'=\cdot$
(\ref{gge}) reduces to (\ref{noinsertion}).
For fundamental lines with $\lambda=\lambda'=\Y(1)$
the decompositions of $\lambda$ and $\lambda'$ giving non-vanishing contributions to (\ref{gge})
are
\begin{align}
(\lambda_0,\lambda_x,\lambda_y)=(\lambda'_0,\lambda'_x,\lambda'_y)
=(\y(1),\cdot,\cdot),(\cdot,\y(1),\cdot),(\cdot,\cdot,\y(1))
\end{align}
and (\ref{gge}) reduces to (\ref{ggefund}).

\paragraph{Leading corrections}

The leading finite $N$ corrections in
line operator indices were
studied in \cite{Beccaria:2024oif}.
Let us confirm $S_{N,\lambda,\lambda'}^{(1,0)}$ obtained by setting $m_x=1$ and $m_y=0$ in
(\ref{gge}) correctly reproduces
the results in \cite{Beccaria:2024oif}.

Because $F_{0,\lambda,\lambda'}^{(y)}=0$ except for $F_{0,\cdot,\cdot}^{(y)}=F_0^{(y)}=1$,
only decompositions with $\lambda_y=\lambda_y'=\cdot$ contribute
to $S_{N,\lambda,\lambda'}^{(1,0)}$;
\begin{align}
S_{N,\lambda,\lambda'}^{(1,0)}
&=
(-1)^{\ell(\lambda)+\ell(\lambda')}
x^N\sigma_xI_1
\nonumber\\&
\times
\sum_{\lambda=\lambda_x+\lambda_0}
\sum_{\lambda'=\lambda'_x+\lambda'_0}
\frac{z_{\lambda_0}}{x^{|\lambda_x|}}
\delta_{\lambda_0,\lambda'_0}
\delta_{|\lambda_x|,|\lambda'_x|}
[I_{\rm F1}]_{\lambda_0}
[I_{\rm F1}^{(x)}]_{\lambda_x+\lambda'_x}.
\label{singlew}
\end{align}
We used the relation
\begin{align}
F^{(x)}_{1,\lambda_x,\lambda_x'}
=\sigma_x I_{1,\lambda_x,\lambda_x'}
=\delta_{|\lambda_x|,|\lambda_x'|}\sigma_xI_1.
\label{fone}
\end{align}
The second equality holds because
in a $U(1)$ theory the insertion
of line operators with $|\lambda|=|\lambda'|$ does not affect the index.
Thanks to (\ref{fone}) the giant graviton index $F_1^{(x)}=\sigma_xI_1$ is factorized in 
(\ref{singlew}) as was pointed out in \cite{Beccaria:2024oif}.
The sum in the second line in
(\ref{singlew}) is a rational function depending on $(\lambda,\lambda')$.
Let us consider two examples studied in
\cite{Beccaria:2024oif}.

The first example is $S_{N,\y(1,1),\y(1,1)}^{(1,0)}$:
\begin{align}
S_{N,\y(1,1),\y(1,1)}^{(1,0)}
&=x^N\sigma_xI_1
\left(2I_{\rm F1}^2
+\frac{4}{x}I_{\rm F1}(I_{\rm F1}^{(x)})^2
+\frac{1}{x^2}(I_{\rm F1}^{(x)})^4
\right).
\end{align}
The three terms in the parentheses correspond to
$(\lambda_0,\lambda_x;\lambda_0',\lambda_x')
=(\y(1,1),\cdot;\y(1,1),\cdot)$,
$(\y(1),\y(1);\y(1),\y(1))$,
and
$(\cdot,\y(1,1);\cdot,\y(1,1))$,
respectively.
The ratio to the corresponding large $N$ index
$I_{\infty,\y(1,1),\y(1,1)}/I_{\rm sugra}=2I_{\rm F1}^2$
is
\begin{align}
\frac{I_{N,\y(1,1),\y(1,1)}^{(1,0)}}{I_{\infty,\y(1,1),\y(1,1)}}
&=x^N\sigma_xI_1
\left(1
+\frac{2}{x}
\frac{(I_{\rm F1}^{(x)})^2}{I_{\rm F1}}
+\frac{1}{2x^2}
\frac{(I_{\rm F1}^{(x)})^4}{I_{\rm F1}^2}
\right)
\nonumber\\
&=x^N\sigma_xI_1
\left(1
+\frac{1+3x-5xy+x^2y}{2x^2}\frac{(1-x)(1-xy)}{(1-y)^2}
\right),
\end{align}
where we introduced the notation
$I_{N,\lambda,\lambda'}^{(m_x,m_y)}=I_{\rm sugra}S_{N,\lambda,\lambda'}^{(m_x,m_y)}$.
This correctly reproduces the corresponding result in \cite{Beccaria:2024oif}.

The second example is
$S_{N,\{Q\},\{Q\}}^{(1,0)}$ with an arbitrary positive integer $Q$.
(\ref{singlew}) gives
\begin{align}
S_{N,\{Q\},\{Q\}}^{(1,0)}
&=x^N\sigma_xI_1
\left(
Q[I_{\rm F1}]_{\{Q\}}
+
\frac{1}{x^Q}
([I_{\rm F1}^{(x)}]_{\{Q\}})^2
\right).
\end{align}
The two terms in the parentheses correspond to
$(\lambda_0,\lambda_x;\lambda_0',\lambda_x')
=(\{Q\},\cdot;\{Q\},\cdot)$
and
$(\cdot,\{Q\};\cdot,\{Q\})$, respectively.
The ratio to the corresponding large $N$ index
$I_{\infty,\{Q\},\{Q\}}/I_{\rm sugra}=Q[I_{\rm F1}]_{\{Q\}}$
is
\begin{align}
\frac{I_{N,\{Q\},\{Q\}}^{(1,0)}}{I_{\infty,\{Q\},\{Q\}}}
&=x^N\sigma_xI_1
\left(
1+
\frac{1}{Qx^Q}
\frac{([I_{\rm F1}^{(x)}]_{\{Q\}})^2}{[I_{\rm F1}]_{\{Q\}}}
\right)
\nonumber\\
&=x^N\sigma_xI_1
\left(
1+
\frac{(1-x^Q)(1-x^Qy^Q)}{Qx^Q(1-y^Q)}
\right) .
\end{align}
Again, this agrees with the corresponding result in \cite{Beccaria:2024oif}.

%%%%%%%%%%%%%%%%%%%%%%%%%%%%%%%%%%%%%%
\paragraph{Numerical test}
Let us numerically confirm that (\ref{gge}) reproduces
the correct indices for different $\lambda$, $\lambda'$, and $N$.

We define $\Delta_m$ as the difference of the line operator index divided by the supergravity index
and the sum over wrapping numbers with the cut-off $m_x+m_y<m$:
\begin{align}
\Delta_m=\frac{I_{N,\lambda,\lambda'}}{I_{\rm sugra}}-\sum_{m_x+m_y<m}S^{(m_x,m_y)}_{N,\lambda,\lambda'}.
\label{delatm-def}
\end{align}
We calculate $\Delta_m$ for different $(N,\lambda,\lambda')$.
As an example, we first consider the case with $(N,\lambda,\lambda')=(3,\y(2,1),\y(1,1,1))$ in detail,
and then we will summarize the results for other cases with
different $(N,\lambda,\lambda')$.

The index for $(N,\lambda,\lambda')=(3,\y(2,1),\y(1,1,1))$
calculated on the gauge theory side by the localization
formula gives the ratio
\begin{align}
\frac{I_{3,\y(2,1),\y(1,1,1)}}{I_{\rm sugra}}
&=2t+13t^2+22t^3-44t^5-62t^6-46t^7-39t^8+16t^9+118t^{10}+{\cal O}(t^{\geq11}) ,
\label{igauge}
\end{align}
where we only show the unrefined results with $x=y=t$ for readability.
(The analysis was done without the unrefinement.)
We use the notation ${\cal O}(t^{\geq n})$ to indicate
the ${\cal O}(t^n)$ term may be zero and the precise order may be
higher than $n$.
We want to confirm this is reproduced
by summing up
$S_{3,\y(2,1),\y(1,1,1)}^{(m_x,m_y)}$ as in (\ref{gge0}).
For each pair of wrapping numbers $(m_x,m_y)$
the six decompositions
shown in Table \ref{decompositions}
contribute to
$S^{(m_x,m_y)}_{N,\y(2,1),\y(1,1,1)}$;
\begin{align}
S^{(m_x,m_y)}_{N,\y(2,1),\y(1,1,1)}
=&x^{Nm_x}y^{Nm_y}(xy)^{m_xm_y}
\nonumber\\&
\times\bigg[
-\frac{1}{y^3}F^{(x)}_{m_x,\cdot,\cdot}F^{(y)}_{m_y,\y(2,1),\y(1,1,1)}[I_{\rm F1}^{(y)}]_{\y(2,1,1,1,1)}
-\frac{3}{x^2y}F^{(x)}_{m_x,\y(2),\y(1,1)}F^{(y)}_{m_y,\y(1),\y(1)}[I_{\rm F1}^{(x)}]_{\y(2,1,1)}[I_{\rm F1}^{(y)}]_{\y(1,1)}
\nonumber\\&
-\frac{3}{xy^2}F^{(x)}_{m_x,\y(1),\y(1)}F^{(y)}_{m_y,\y(2),\y(1,1)}[I_{\rm F1}^{(x)}]_{\y(1,1)} [I_{\rm F1}^{(y)}]_{\y(2,1,1)}
-\frac{3}{y^2}I_{\rm F1}F^{(x)}_{m_x,\cdot,\cdot}F^{(y)}_{m_y,\y(2),\y(1,1)} [I_{\rm F1}^{(y)}]_{\y(2,1,1)}
\nonumber\\&
-\frac{1}{x^3}F^{(x)}_{m_x,\y(2,1),\y(1,1,1)}F^{(y)}_{m_y,\cdot,\cdot} [I_{\rm F1}^{(x)}]_{\y(2,1,1,1,1)}
-\frac{3}{x^2}I_{\rm F1}F^{(x)}_{m_x,\y(2),\y(1,1)}F^{(y)}_{m_y,\cdot,\cdot}[I_{\rm F1}^{(x)}]_{\y(2,1,1)}
\Bigg] .
\end{align}
The contributions from small values of the total wrapping number $m=m_x+m_y$ are
\begin{align}
S^{(0,0)}_{3,\y(2,1),\y(1,1,1)}&=0,
\nonumber\\
S^{(m_x+m_y=1)}_{3,\y(2,1),\y(1,1,1)}
&=2t+13t^2+22t^3-4t^4-76t^5-193t^6
-344t^8-640t^9
\nonumber\\&\quad
-784t^{10}+{\cal O}(t^{\geq11}),
\nonumber\\
S^{(m_x+m_y=2)}_{3,\y(2,1),\y(1,1,1)}
&=4t^4+32t^5+131t^6+298t^7+457t^8+650t^9
+854t^{10}+{\cal O}(t^{\geq11}),
\nonumber\\
S^{(m_x+m_y=3)}_{3,\y(2,1),\y(1,1,1)}
&=6t^9+48t^{10}+{\cal O}(t^{\geq11}),
\label{s3ll}
\end{align}
where we used the notation
\begin{align}
S^{(m_x+m_y=m)}_{N,\lambda,\lambda'}:=
\sum_{m_x+m_y=m}S^{(m_x,m_y)}_{N,\lambda,\lambda'}.
\end{align}
By subtracting (\ref{s3ll}) from (\ref{igauge}) one by one,
we confirm that as $m$ increases, the order of $\Delta_m$ increases;
\begin{align}
\Delta_1={\cal O}(t^1),\quad
\Delta_2={\cal O}(t^4),\quad
\Delta_3={\cal O}(t^9),\quad
\Delta_4={\cal O}(t^{\geq11}).
\end{align}

\newcommand{\sh}{\cellcolor[gray]{0.8}}

We carried out the same analysis for $N=1,2,3,4$, $m\geq4$ and $|\lambda|=|\lambda'|\leq4$.
See Table \ref{texpresults} for extracted data from the results.
\begin{table}[htb]
\caption{The orders of discrepancies $\Delta_m$.
$\Delta_0$ is the ratio $I_{N,\lambda,\lambda'}/I_{\rm sugra}$ itself without subtraction.
In most cases the order saturates the bound in (\ref{deltam}),
while in some cases, which are shaded in the table, the coefficient of the leading term accidentally
vanishes and the order becomes higher.
In calculation of $\Delta_4$ only the first two terms in $F_3^{(x/y)}$ are used,
and we did not determine the precise order of the leading term.}\label{texpresults}
\centering
\begin{tabular}{cccccc}
\hline
\hline
                  & $\Delta_0$ & $\Delta_1$ & $\Delta_2$ & $\Delta_3$ & $\Delta_4$ \\
\hline
$I_{1,\y(2),\y(2)}$ & ${\cal O}(t^0)$ & ${\cal O}(t^0)$ & ${\cal O}(t^2)$ & ${\cal O}(t^6)$ & \sh ${\cal O}(t^{\geq8})$ \\
$I_{2,\y(2),\y(2)}$ & ${\cal O}(t^0)$ & ${\cal O}(t^1)$ & ${\cal O}(t^4)$ & ${\cal O}(t^9)$ & \sh ${\cal O}(t^{\geq11})$ \\
$I_{3,\y(2),\y(2)}$ & ${\cal O}(t^0)$ & ${\cal O}(t^2)$ & ${\cal O}(t^6)$ & ${\cal O}(t^{12})$ & \sh ${\cal O}(t^{\geq14})$ \\
$I_{4,\y(2),\y(2)}$ & ${\cal O}(t^0)$ & ${\cal O}(t^3)$ & ${\cal O}(t^8)$ & ${\cal O}(t^{15})$ & \sh ${\cal O}(t^{\geq17})$ \\
\hline
$I_{1,\y(2,1),\y(1,1,1)}$ & ${\cal O}(t^0)$ &  \sh${\cal O}(t^0)$ & ${\cal O}(t^0)$ & ${\cal O}(t^3)$ & \sh${\cal O}(t^{\geq5})$ \\
$I_{2,\y(2,1),\y(1,1,1)}$ & ${\cal O}(t^0)$ & ${\cal O}(t^0)$ & ${\cal O}(t^2)$ & ${\cal O}(t^6)$ & \sh${\cal O}(t^{\geq8})$ \\
$I_{3,\y(2,1),\y(1,1,1)}$ & \sh${\cal O}(t^1)$ & ${\cal O}(t^1)$ & ${\cal O}(t^4)$ & ${\cal O}(t^9)$ & \sh${\cal O}(t^{\geq11})$ \\
$I_{4,\y(2,1),\y(1,1,1)}$ & \sh${\cal O}(t^2)$ & ${\cal O}(t^2)$ & ${\cal O}(t^6)$ & ${\cal O}(t^{12})$ & \sh${\cal O}(t^{\geq14})$ \\
\hline
$I_{1,\y(3,1),\y(2,2)}$ & ${\cal O}(t^0)$ & \sh${\cal O}(t^0)$ & ${\cal O}(t^{-2})$ & ${\cal O}(t^0)$ & \sh${\cal O}(t^{\geq2})$ \\
$I_{2,\y(3,1),\y(2,2)}$ & ${\cal O}(t^0)$ & \sh${\cal O}(t^0)$ & ${\cal O}(t^0)$ & ${\cal O}(t^3)$ & \sh${\cal O}(t^{\geq5})$ \\
$I_{3,\y(3,1),\y(2,2)}$ & ${\cal O}(t^0)$ & ${\cal O}(t^0)$ & ${\cal O}(t^2)$ & ${\cal O}(t^6)$ & \sh${\cal O}(t^{\geq8})$ \\
$I_{4,\y(3,1),\y(2,2)}$ & \sh${\cal O}(t^1)$ & ${\cal O}(t^1)$ & ${\cal O}(t^4)$ & ${\cal O}(t^9)$ & \sh${\cal O}(t^{\geq11})$ \\
\hline
\end{tabular}
\end{table}
We found the order of $\Delta_m$ is bounded by
\begin{align}
\Delta_m={\cal O}(t^{\geq m(N-k)+m^2}) , \label{deltam}
\end{align}
where $k=|\lambda|=|\lambda'|$.
If we accept this,
we can conclude that the correct index
is obtained in the $m\rightarrow\infty$ limit.

\section{Simple-sum expansion}\label{simplesum.sec}
It was found in \cite{Gaiotto:2021xce} that
if we treat the index as a Taylor series in the fugacity $y$,
contributions from $m_y\geq1$ become trivial,
and the double-sum expansion (\ref{gge})
reduces to the simple-sum expansion.
See \cite{Imamura:2022aua} and \cite{Fujiwara:2023bdc}
for some discussions on the reduction.
Let us confirm that it is the case for the line-operator indices
$I_{N,\lambda,\lambda'}$.
Namely, we want to confirm that the simple-sum expansion
\begin{align}
\frac{I_{N,\lambda,\lambda'}}{I_{\rm sugra}}
&=
\sum_{m_x=0}^\infty S_{N,\lambda,\lambda'}^{(m_x,0)}
\label{reducedgge}
\end{align}
holds.
$S_{N,\lambda,\lambda'}^{(m_x,0)}$
are obtained by setting $m_y=0$ in (\ref{gge}).
Because $F_{0,\lambda_y,\lambda_y}^{(y)}$ vanishes except for $F_{0,\cdot,\cdot}^{(y)}=F_0^{(y)}=1$,
only decompositions of $(\lambda,\lambda')$ with $\lambda_y=\lambda_y'=\cdot$
gives a non-vanishing contribution to $S_{N,\lambda,\lambda'}^{(m_x,0)}$,
and we obtain
\begin{align}
S_{N,\lambda,\lambda'}^{(m_x,0)}&=
(-1)^{\ell(\lambda)+\ell(\lambda')}
\sum_{\lambda=\lambda_x+\lambda_0}
\sum_{\lambda'=\lambda'_x+\lambda'_0}
z_{\lambda_0}\delta_{\lambda_0,\lambda'_0}[I_{\rm F1}]_{\lambda_0}
\nonumber\\&
\times x^{m_xN}\frac{1}{x^{|\lambda_x|}}F^{(x)}_{m_x,\lambda_x,\lambda'_x}[I_{\rm F1}^{(x)}]_{\lambda_x+\lambda'_x}.
\label{ssimple}
\end{align}

Let us numerically
confirm (\ref{reducedgge}) holds.
We first consider an example with $(N\,\lambda,\lambda')=(3,\y(2,1),\y(1,1,1))$
in detail, and 
then we will summarize the analysis for different $(N\,\lambda,\lambda')$.
On the gauge theory side we obtain
\begin{align}
I_{3,\y(2,1),\y(1,1,1)}/I_{\rm sugra}
&=(x+4x^2+6x^3+6x^4+2x^5-4x^6-11x^7-17x^8+\cdots)
\nonumber\\
&+(1+5x+5x^2-3x^3-12x^4-17x^5-12x^6-3x^7+9x^8+\cdots)y
\nonumber\\
&+(4+5x-6x^2-12x^3-8x^4+2x^5+5x^6+3x^7+x^8+\cdots)y^2
\nonumber\\
&+{\cal O}(y^3).
\label{i3llxy}
\end{align}
(In the unrefined limit $x,y\rightarrow t$ this becomes (\ref{igauge}).)
We want to confirm this is reproduced by the
expansion (\ref{reducedgge}).
Let us calculate the discrepancy for
the truncated sum
\begin{align}
\Delta_m=\frac{I_{N,\lambda,\lambda'}}{I_{\rm sugra}}
-\sum_{m_x=0}^{m-1}S_{N,\lambda,\lambda'}^{(m_x)}
\end{align}
depending on the cutoff $m$.
The decompositions of $\lambda=\y(2,1)$ and $\lambda'=\y(1,1,1)$ contributing to
$S^{(m_x,0)}_{N,\y(2,1),\y(1,1,1)}$ are shown in Table \ref{dec2}.
\begin{table}[htb]
\caption{
List of decompositions giving non-trivial contributions for
$\lambda=\protect\y(2,1)$ and
$\lambda'=\protect\y(1,1,1)$.
The third column shows the multiplicities.}\label{dec2}
\centering
\begin{tabular}{ccccc}
\hline
\hline
$(\lambda_0,\lambda_x)$ & $(\lambda'_0,\lambda'_x)$ & mult. \\
\hline
$(\cdot,\y(2,1))$ & $(\cdot,\y(1,1,1))$ & $1$ \\
$(\y(1),\y(2))$ & $(\y(1),\y(1,1))$ & $3$ \\
\hline
\end{tabular}
\end{table}
Corresponding to the two rows in the table
$S^{(m_x,0)}_{N,\y(2,1),\y(1,1,1)}$ is given as the sum of two terms;
\begin{align}
S^{(m_x,0)}_{N,\y(2,1),\y(1,1,1)}&=
x^{m_xN}
\left[
-\frac{1}{x^3}F^{(x)}_{m_x,\y(2,1),\y(1,1,1)}[I_{\rm F1}^{(x)}]_{\y(2,1,1,1,1)}
-\frac{3}{x^2}I_{\rm F1}F^{(x)}_{m_x,\y(2),\y(1,1)}[I_{\rm F1}^{(x)}]_{\y(2,1,1)}
\right].
\end{align}
$S^{(m_x,0)}_{3,\y(2,1),\y(1,1,1)}$
with $m_x=0,1,2,3$ are
given explicitly as follows:
\begin{align}
S_{3,\y(2,1),\y(1,1,1)}^{(0,0)}
&=0,
\nonumber\\
%%%%%%%%%%%%%%%%%%%%%%%%%%%%%%%%%%%%
S_{3,\y(2,1),\y(1,1,1)}^{(1,0)}
&=(x+4x^2+\cdots)
+(1+6x+\cdots)y
+(x^{-1}+7+\cdots) y^2
+\cdots,\nonumber\\
%%%%%%%%%%%%%%%%%%%%%%%%%%%%%%%%%%%%
S_{3,\y(2,1),\y(1,1,1)}^{(2,0)}
&=(-x^3-4x^4+\cdots)
+(-x-4x^2+\cdots)y
+(-x^{-1}-4+\cdots)y^2
+\cdots,\nonumber\\
%%%%%%%%%%%%%%%%%%%%%%%%%%%%%%%%%%%%
S_{3,\y(2,1),\y(1,1,1)}^{(3,0)}
&=(x^6+4x^7+\cdots)
+(x^3+4x^4+\cdots)y
+(1+4x+\cdots)y^2
+\cdots.
\end{align}
We see that by subtracting these
from (\ref{i3llxy}) one by one
low order terms
in (\ref{i3llxy}) are gradually canceled.
\begin{align}
\Delta_1
&=(x+4x^2+\cdots)
+(1+5x+\cdots)y
+(4+5x+\cdots)y^2+\cdots,\nonumber\\
\Delta_2
&=(-x^3-4x^4+\cdots)
+(-x+4x^2+\cdots)y
+(-x^{-1}-3+\cdots)y^2+\cdots,\nonumber\\
\Delta_3
&=(x^6+4x^7+\cdots)
+(x^3+4x^4+\cdots)y
+(1+4x+\cdots)y^2+\cdots,\nonumber\\
\Delta_4
&=(-x^{10}-4x^{11}+\cdots)
+(-x^6-4x^7+\cdots)y
+(-x^2-4x^3+\cdots)y^2
+\cdots.
\end{align}

We carried out the same  analysis for different $(N,\lambda,\lambda')$.
Some data extracted from the results are shown in Table \ref{ysrdata}.
\begin{table}[htb]
\caption{The orders of the first few coefficients of $y$-expansion of $\Delta_m$
are shown for a few examples of $(N,\lambda,\lambda')$.
$\Delta_0$ is the ratio $I_{N,\lambda,\lambda'}/I_{\rm sugra}$ itself without subtraction.
Most of the coefficients saturate the bound
in (\ref{ysrbound}), while some of them, which are shaded, have higher order.
}\label{ysrdata}
\centering
\begin{tabular}{cc@{\ }c@{\ }c@{\ }c@{\ }c}
\hline
\hline
$I_{3,\y(2),\y(2)}$
& $\Delta_0=$ & ${\cal O}(x^0)y^0$&\sh$+{\cal O}(x^1)y^1$&$+{\cal O}(x^0)y^2$&$+{\cal O}(y^3)$ \\
& $\Delta_1=$ & ${\cal O}(x^2)y^0$&$+{\cal O}(x^1)y^1$&$+{\cal O}(x^0)y^2$&$+{\cal O}(y^3)$ \\
& $\Delta_2=$ & ${\cal O}(x^5)y^0$&$+{\cal O}(x^3)y^1$&$+{\cal O}(x^1)y^2$&$+{\cal O}(y^3)$ \\
& $\Delta_3=$ & ${\cal O}(x^9)y^0$&$+{\cal O}(x^6)y^1$&$+{\cal O}(x^3)y^2$&$+{\cal O}(y^3)$ \\
& $\Delta_4=$ & ${\cal O}(x^{14})y^0$&$+{\cal O}(x^{10})y^1$&$+{\cal O}(x^6)y^2$&$+{\cal O}(y^3)$ \\
\hline
$I_{3,\y(2,1),\y(1,1,1)}$
& $\Delta_0=$ & $\sh{\cal O}(x^1)y^0$&$+{\cal O}(x^0)y^1$&$+{\cal O}(x^0)y^2$&$+{\cal O}(y^3)$ \\
& $\Delta_1=$ & ${\cal O}(x^1)y^0$&$+{\cal O}(x^0)y^1$&$\sh+{\cal O}(x^0)y^2$&$+{\cal O}(y^3)$ \\
& $\Delta_2=$ & ${\cal O}(x^3)y^0$&$+{\cal O}(x^1)y^1$&$+{\cal O}(x^{-1})y^2$&$+{\cal O}(y^3)$ \\
& $\Delta_3=$ & ${\cal O}(x^6)y^0$&$+{\cal O}(x^3)y^1$&$+{\cal O}(x^0)y^2$&$+{\cal O}(y^3)$ \\
& $\Delta_4=$ & ${\cal O}(x^{10})y^0$&$+{\cal O}(x^6)y^1$&$+{\cal O}(x^2)y^2$&$+{\cal O}(y^3)$ \\
\hline
$I_{4,\y(3,1),\y(2,2)}$
& $\Delta_0=$ & $\sh{\cal O}(x^1)y^0$&$+{\cal O}(x^0)y^1$&$+{\cal O}(x^0)y^2$&$+{\cal O}(y^3)$ \\
& $\Delta_1=$ & ${\cal O}(x^1)y^0$&$+{\cal O}(x^0)y^1$&$\sh+{\cal O}(x^0)y^2$&$+{\cal O}(y^3)$ \\
& $\Delta_2=$ & ${\cal O}(x^3)y^0$&$+{\cal O}(x^1)y^1$&$+{\cal O}(x^{-1})y^2$&$+{\cal O}(y^3)$ \\
& $\Delta_3=$ & ${\cal O}(x^6)y^0$&$+{\cal O}(x^3)y^1$&$+{\cal O}(x^0)y^2$&$+{\cal O}(y^3)$ \\
& $\Delta_4=$ & ${\cal O}(x^{10})y^0$&$+{\cal O}(x^6)y^1$&$+{\cal O}(x^2)y^2$&$+{\cal O}(y^3)$ \\
\hline
\end{tabular}
\end{table}

The numerical analysis shows the bound
\begin{align}
\Delta_m
&=\sum_{i=0}^\infty{\cal O}\left(x^{\geq\frac{1}{2}m(m+1)+m(N-k-i)}\right)y^i .
\label{ysrbound}
\end{align}
We have confirmed this holds for $m\leq4$, $N=1,2,3$, $|\lambda|\leq3$, and $i\leq 3$.
If (\ref{ysrbound}) always holds, the index is correctly reproduced in the $m\rightarrow\infty$ limit.

%%%%%%%%%%%%%%%%%%%%%%%%%%%%%%%%%%%%%%%%%%%%%%
\section{Conclusions and Discussion}
As a natural generalization of large $N$ line-operator indices
in arbitrary representations (\ref{largen})
and the giant graviton expansion for the fundamental line-operator index (\ref{ggefund}),
we proposed a double-sum giant graviton expansion of Wilson
line-operator indices in arbitrary representations (\ref{gge0}) and (\ref{gge}).
We adopted the holographic realization of line operators by fundamental strings, and
a weight $k$ representation is realized by $k$ coincident fundamental strings.
In this picture, it is natural to use the symmetric power sum polynomials $p_\lambda$
rather than the Schur polynomials
as the basis of symmetric polynomials, and a partition $\lambda$ specifies not an irreducible $U(N)$ representation
but the Chan-Paton structure of the worldsheet around the temporal circle.
Coincident worldsheets consist of $\ell(\lambda)$ disconnected components,
and each component may or may not end on giant gravitons.
There are different ways of connecting semi-infinite strings,
and by summing up all contributions
from them, we obtain the finite $N$ line operator index.
The formula (\ref{gge0}) with (\ref{gge}) passes highly non-trivial numerical tests.
We also numerically confirmed the simple-sum version
of the expansion (\ref{reducedgge}) and (\ref{ssimple})
reproduces correct indices.

There are many ways of extending our results.
One is to consider composite systems of F1, D5, and D3-branes extending along $AdS_2$.
The F1 description of line operators in this paper is appropriate
for representations with weights  of order one.
For the symmetric and anti-symmetric representations with weight of order $N$,
it is more natural to realize the line operators by D3-branes
\cite{Drukker:2005kx} and D5-branes \cite{Yamaguchi:2006tq}, respectively.
In particular, the line-operator index for anti-symmetric representations
is reproduced on the AdS side by D5-brane for large $N$ \cite{Gang:2012yr}
and for finite $N$ \cite{Imamura:2024pgp}.
We can combine them to realize more complicated line operators corresponding
to a Young diagram consisting of a few long columns, a few long rows, and a few boxes.
Naive expectation is that we can realize such a line operator
as a composite system of F1, D3, and D5-branes.
It is interesting to investigate giant graviton expansions for such line operators.

For representation with weight of order $N^2$,
the line operators are holographically realized by bubbling geometries
\cite{DHoker:2007mci}, and indices for such line operators
are studied in \cite{Hatsuda:2023imp}.
It is interesting to consider the finite $N$ correction in such cases.

It is also interesting to consider other theories.
In this work, we investigated ${\cal N}=4$ $U(N)$ SYM exclusively.
We can also consider ${\cal N}=4$ SYM with other gauge groups and
more general  field theories with fewer supersymmetries.
We hope we will return to these problems near future.

\section*{Acknowledgments}
The authors thank M.~Inoue for valuable discussions.
The work of Y.~I. and D.~Y. was supported by JSPS KAKENHI Grant Number JP21K03569, 
and the work of A.~S was supported by JSPS KAKENHI Grant Number JP24KJ1105.

\appendix
\section{Giant graviton indices $F^{(x)}_{m,\lambda,\lambda'}$}\label{app.sec}
In this appendix, we show explicit forms of $F^{(x)}_{m,\lambda,\lambda'}$
for small $m$, $\lambda$, and $\lambda'$.

For $m=0$, the functions vanish unless the partitions are empty:
\begin{align}
F_{0,\lambda,\lambda'}^{(x)}=0\quad
\mbox{except for}\quad
F_{0,\cdot,\cdot}^{(x)}=1.
\end{align}

The functions $F_{1,\lambda,\lambda'}^{(x)}$
agree with $F_1^{(x)}$ if $|\lambda|=|\lambda'|$;
\begin{align}
F_{1,\lambda,\lambda'}^{(x)}=F_1^{(x)}
\delta_{|\lambda|,|\lambda'|}
=\delta_{|\lambda|,|\lambda'|}\Pexp(\sigma_xf^{({\rm vec})}) .
\end{align}
In the analysis of the double-sum expansion in Section \ref{doublesum.sec}
we treat the index as a function of $t=\sqrt{xy}$ and $v=\sqrt{y/x}$,
and use $t$-series expansion in numerical analysis.
The first few terms in the $t$-series expansion of $F_1^{(x)}$ are
\begin{align}
F_1^{(x)}=\frac{-x}{1-\frac{y}{x}}+x^2(-1+\tfrac{y}{x})+{\cal O}(t^3).
\end{align}
In the analysis of the simple-sum expansion in Section \ref{simplesum.sec}
we use $y$-series expansion in numerical analysis.
The first few terms in the $y$-series expansion of $F_1^{(x)}$ are
\begin{align}
F_1^{(x)}=\frac{-x}{1-x}+(-1+x)y+{\cal O}(y^2) .
\end{align}

The functions $F_{m,\lambda,\lambda'}^{(x)}$ with $m\geq2$
can be obtained by holonomy integrals similar to (\ref{localization})
with $f^{({\rm vec})}$ replaced by $\sigma_xf^{({\rm vec})}$.
We need to choose the integration contours carefully.
See \cite{Arai:2020qaj,Imamura:2021ytr} for some explanations
about the contours.
In the following, we show the results for $m=2,3$ with $|\lambda|=|\lambda'|\leq2$.

The $t$-series expansions of $F_{m,\lambda,\lambda'}$ are
\begin{align}
F_{2,\cdot,\cdot}^{(x)}
&=\frac{\frac{x^5}{y}(1-2\frac{y}{x})}{(1-\frac{y}{x})(1-\frac{y^2}{x^2})}
+x^5(2-\tfrac{x^2}{y^2})+{\cal O}(t^6),\nonumber\\
%%%%%%%
F_{2,\y(1),\y(1)}^{(x)}
&=\frac{-\frac{x^4}{y}}{(1-\frac{y^2}{x^2})}
+\frac{-\frac{x^6}{y^2}(1+\frac{y}{x}-5\frac{y^2}{x^2}+\frac{y^3}{x^3}+\frac{y^4}{x^4}-\frac{y^5}{x^5})}{(1-\frac{y}{x})(1-\frac{y^2}{x^2})}+{\cal O}(t^5),\nonumber\\
%%%%%%%
F_{2,\y(2),\y(2)}^{(x)}
&=\frac{-\frac{x^3}{y}}{(1-\frac{y^2}{x^2})}
+x^3(1-\tfrac{x^2}{y^2})+{\cal O}(t^4),\nonumber\\
F_{2,\y(2),\y(1,1)}^{(x)}
&=\frac{-\frac{x^3}{y}}{(1-\frac{y^2}{x^2})}
+\frac{-\frac{x^5}{y^2}(1+2\frac{y}{x}-2\frac{y^2}{x^2}+\frac{y^4}{x^4})}{(1-\frac{y^2}{x^2})}+{\cal O}(t^4),\nonumber\\
F_{2,\y(1,1),\y(1,1)}^{(x)}
&=\frac{-\frac{x^3}{y}}{(1-\frac{y^2}{x^2})}
+\frac{-\frac{x^5}{y^2}(1+4\frac{y}{x}-2\frac{y^2}{x^2}+\frac{y^4}{x^4})}{(1-\frac{y^2}{x^2})}+{\cal O}(t^4),\nonumber\\
%%%%%%%%%%%%%%%%%%%
F_{3,\cdot,\cdot}^{(x)}
&=\frac{\frac{x^{12}}{y^3}(-2+3\frac{y}{x}+3\frac{y^2}{x^2}-5\frac{y^3}{x^3})}{(1-\frac{y}{x})(1-\frac{y^2}{x^2})(1-\frac{y^3}{x^3})}
+\frac{\frac{x^{15}}{y^5}(-2+3\frac{y^2}{x^2}+3\frac{y^3}{x^3}-5\frac{y^5}{x^5})}{1-\frac{y^2}{x^2}}+{\cal O}(t^{11}),
\nonumber\\
F^{(x)}_{3,\y(1),\y(1)}
&=\frac{-\frac{x^{10}}{y^3}}{1-\frac{y^3}{x^3}}
+\frac{\frac{x^{13}}{y^5}(-1-2\frac{y^2}{x^2}+\frac{y^3}{x^3}+5\frac{y^4}{x^4}-\frac{y^6}{x^6}+\frac{y^8}{x^8})}{(1-\frac{y^2}{x^2})(1-\frac{y^3}{x^3})}
+{\cal O}(t^9),
\nonumber\\
%%%%%%%%%%%%%%%%%%%%%%%%%%%%%%%%%%%
F^{(x)}_{3,\y(2),\y(2)}
&=\frac{-\frac{x^8}{y^3}}{1-\frac{y^3}{x^3}}
+\tfrac{x^{11}}{y^5}(-1+\tfrac{y^3}{x^3})
+{\cal O}(t^7),\nonumber\\
%%%%%%%%%%%%%%%%%%%%%%%%%%%%%%%%%%%
F^{(x)}_{3,\y(2),\y(1,1)}
&=\frac{-\frac{x^8}{y^3}}{1-\frac{y^3}{x^3}}
+\frac{\frac{x^{11}}{y^5}(-1-2\frac{y^2}{x^2}+2\frac{y^3}{x^3}-\frac{y^6}{x^6})}{1-\frac{y^3}{x^3}}
+{\cal O}(t^7),\nonumber\\
%%%%%%%%%%%%%%%%%%%%%%%%%%%%%%%%%%%
F^{(x)}_{3,\y(1,1),\y(1,1)}
&=\frac{-\frac{x^8}{y^3}}{1-\frac{y^3}{x^3}}
+\frac{\frac{x^{11}}{y^5}(-1-4\frac{y^2}{x^2}+2\frac{y^3}{x^3}-\frac{y^6}{x^6})}{1-\frac{y^3}{x^3}}
+{\cal O}(t^7).
\end{align}

The $y$-series expansions of $F_{m,\lambda,\lambda'}$ are
\begin{align}
F_{2,\cdot,\cdot}^{(x)}
&=\frac{x^3}{(1-x)(1-x^2)}+xy+{\cal O}(y^2),\nonumber\\
%%%%%%%%%%%%%%%%%%%%%%%%
F_{2,\y(1),\y(1)}^{(x)}
&=\frac{x^2}{(1-x)^2}+(1+2x)y+{\cal O}(y^2),\nonumber\\
%%%%%%%%%%%%%%%%%%%%%%%%
F_{2,\y(2),\y(2)}^{(x)}
&=\frac{x(1-x+2x^2)}{(1-x)(1-x^2)}+(\tfrac{1}{x}+x)y+{\cal O}(y^2),\nonumber\\
%%%%%%%%%%%%%%%%%%%%%%%%
F_{2,\y(2),\y(1,1)}^{(x)}
&=\frac{x}{(1-x)(1-x^2)}+\tfrac{1}{x}(1+2x+x^2)y+{\cal O}(y^2),\nonumber\\
%%%%%%%%%%%%%%%%%%%%%%%%
F_{2,\y(1,1),\y(1,1)}^{(x)}
&=\frac{x(1+2x)}{(1-x)^2}+\tfrac{1}{x}(1+4x+5x^2)y+{\cal O}(y^2),\nonumber\\
%%%%%%%%%%%%%%%%%%%%%%%%%
F_{3,\cdot,\cdot}^{(x)}
&=\frac{-x^6}{(1-x)(1-x^2)(1-x^3)}
+\frac{-x^3}{1-x^2}y
+{\cal O}(y^2),
\nonumber\\
F^{(x)}_{3,\y(1),\y(1)}
&=\frac{-x^4}{(1-x)^2(1-x^2)}
+\frac{-x(1+2x+2x^2)}{1-x^2}y
+{\cal O}(y^2),
\nonumber\\
F^{(x)}_{3,\y(2),\y(2)}
&=\frac{-x^2(1-2 x+2 x^2)}{(1-x)^2(1-x^2)}
+\frac{-\frac{1}{x}(1+x^2-x^3+2 x^4)}{1-x^2}y
+{\cal O}(y^2),
\nonumber\\
F^{(x)}_{3,\y(2),\y(1,1)}
&=\frac{-x^2}{(1-x)^2(1-x^2)}
+\frac{-\frac{1}{x}(1+2 x+3 x^2+x^3)}{1-x^2}y
+{\cal O}(y^2),
\nonumber\\
F^{(x)}_{3,\y(1,1),\y(1,1)}
&=\frac{-x^2(1+2 x+2 x^2)}{(1-x)^2(1-x^2)}
+\frac{-\frac{1}{x}(1+4 x+9 x^2+11 x^3+6 x^4)}{1-x^2}y
+{\cal O}(y^{2}).
\end{align}

\end{document}